\font\subtit=cmr12
\font\name=cmr8

\def\plb#1#2#3#4{#1, {\it Phys. Lett.} {\bf {#2}}B (#3), #4}
\def\npb#1#2#3#4{#1, {\it Nucl. Phys.} {\bf B{#2}} (#3), #4}

\def\cmp#1#2#3#4{#1, {\it Comm. Math. Phys.} {\bf {#2}} (#3), #4}

\def\ijmpa#1#2#3#4{#1, {\it Int. Jour. Mod. Phys.} {\bf A{#2}} (#3), #4}
\def\jmp#1#2#3#4{#1, {\it Jour. Math. Phys.} {\bf {#2}} (#3), #4}
\def\hpa#1#2#3#4{#1, {\it Helv. Phys. Acta} {\bf {#2}} (#3), #4}

\def\jgp#1#2#3#4{#1, {\it Journal Geom. Phys.} {\bf {#2}} (#3), #4}

\def\fph#1#2#3#4{#1, {\it Fortschr. Physik} {\bf {#2}} (#3), #4}
\input harvmac
\pageno=0

\null
\hskip 8.2 cm {Preprint PAR-LPTHE 97-38}
\vskip 2truecm
\centerline{\subtit
MONODROMY PROPERTIES OF ENERGY MOMENTUM TENSOR
}
\centerline{\subtit ON GENERAL ALGEBRAIC CURVES
}
\vskip 1truecm
\centerline{F{\name RANCO} F{\name ERRARI}$^{a}$ and J{\name AN} T.
S{\name OBCZYK}$^b$}
\smallskip $^a${\it
LPTHE, Universit\'e Pierre and Maria Curie--Paris VI and
Universit\'e Denis Diderot--Paris VII, Boite 126, Tour 16, 1$^{er}$ \'etage,
4 place Jussieu, F-75252 Paris Cedex 05, France, E-mail:
fferrari@lpthe.jussieu.fr}
\smallskip $^b${\it
Institute for Theoretical Physics, Wroc\l aw
University, pl.  Maxa Borna 9, 50205 Wroc\l aw, Poland, E-mail:
jsobczyk@proton.ift.uni.wroc.pl}
\smallskip
\vskip 3cm
\centerline{ABSTRACT}
{\narrower \abstractfont
A new approach to analyze the properties of the
energy-momentum tensor $T(z)$ of conformal
field theories on generic Riemann surfaces (RS)
is proposed. $T(z)$ is
decomposed into $N$
components with different
monodromy properties, where $N$ is
the number of branches in the realization of RS as branch
covering over the complex sphere. This decomposition gives rise to
new infinite dimensional Lie algebra which can be viewed as a generalization
of Virasoro algebra containing information about the global properties of
the underlying RS. In the simplest case of hyperelliptic
curves the structure of the algebra is calculated in two ways and its
central extension is explicitly given. The algebra possess an interesting
symmetry with a clear interpretation in the framework of the radial
quantization of CFT's with multivalued fields on the complex sphere.
} 
\Date{September 1997}
\vfill\eject
\newsec {INTRODUCTION}
\vskip 1cm
 
The problem of constructing conformal field theories (CFT's) on higher genus
Riemann surfaces (RS)
arose in the middle of the previous decade
in the context of perturbative string theories, where higher loop
computations require the evaluation of CFT's
on higher genus RS
\ref\pol{
\plb{A. M.  Polyakov}{103}{1981}{207, 211}.
}.
CFT's have been studied in connection with
phase transitions phenomena at criticality, where the physics
becomes scale invariant. The structure of CFT's in $D=2$
dimensions turns out to be particularly rich due to the infinite
dimensional group of local conformal transformations generated by the
Virasoro algebra
\ref\cft{P. Di Francesco, P. Mathieu and D. S\' en\' echal, {\it Conformal
Field Theory}, Springer Verlag, 1996.}.
The presence of this algebra allows the complete solution
of certain conformal models
\ref\bpz{\npb{A. A. Belavin, A. M. Polyakov and A. B. Zamolodchikov}
{241}{1984}{333}.}.
It became very soon clear that the conformal symmetry plays a crucial role
in understanding the structure of string theory. The possible consistent
vacua for string theory can be recognized as CFT's
\ref\fms{
\npb{D. Friedan, E. Martinec and S.
Shenker}{271}{1986}{93}.
}.

In investigating CFT's it has been fruitful to study these theories on
the torus. In particular the idea
of modular invariance has found many applications.
Modular invariance involves left and right moving sectors simultaneously
and
provides severe restrictions on the class of "acceptable" CFT's
\ref\torus{
\npb{J. Cardy}{275[FS17]}{1986}{200}.
}.
The properties of CFT's
on more general RS are perhaps not so relevant to physics,
at least from the present point of view.
Yet it is interesting to consider "new examples"
of CFT's. One can talk about "new examples" in the following sense.
Most of the models discussed so far are free field
theories.
They are however nontrivial due to the interaction
with two dimensional gravity and because of the
topology of the manifolds on which the theories
have been defined
\ref\eo{\npb{T.  Eguchi, H. Ooguri}{282}{1987}{308}}
\ref\zama{
\npb{Al. B. Zamolodchikov}{316}{1989}{573}}
\ref\mart{
\jmp{E. Guardagnini, M. Martellini and M. Mintchev}{31}{1990}{1226}.
}.
They can also be viewed as "new" since they can be treated as
CFT's on complex sphere with multivalued conformal fields.

One of the most interesting approaches to analyze CFT's on higher genus RS
is to define an equivalent multivalued CFT on the complex sphere.
To do this, we notice that a
RS can be represented as a
branch covering of the sphere, i.e. it can be visualized as $N$ copies of the
complex sphere "glued" at some branch points
\ref\grha{
P. Griffiths
and J. Harris, {\it Principles of Algebraic
Geometry}, John Wiley \& Sons, New York 1978
}.
The information about the geometry of the RS is
encoded in its
monodromy properties.
In the language of the multivalued CFT's mentioned above, one can construct 
"twist operators" located at the projections of the branch points on the
sphere, which ``simulate'' the effects of the geometry.
In this way the CFT on the multisheeted RS is claimed to be equivalent to the
theory on the complex sphere supplemented by extra twist field insertions.
More precisely,
since a CFT is usually defined by a set of correlation functions with
required analytic properties, equivalence means here that the correlation
functions on the sphere with twist fields are formally the same as the
correlation
functions on the RS.

A first paper containing similar ideas in a slightly different context
of CFT's on orbifolds is that of Dixon et al.
\ref\orb{
\npb{L. Dixon, D. Friedan, E. Martinec and S. Shenker}{282}{1987}{13}.
}.
Later, several more detail papers followed
\ref\bera{
\ijmpa{M.
A.  Bershadsky and A.  O.  Radul}{2}{1987}{165}
}
\ref\zam{
\npb{Al. B. Zamolodchikov}{285}{1987}{481}.
}
\ref\kniz{\cmp{V. G. Knizhnik}
{112}{1987} {587}.}.
In the case of the $b-c$ systems on $Z_N$ symmetric curves,
a very clear picture was obtained in the language of
bosonisation: the twist operators have been explicitly constructed.
The
treatment of bosonic fields with quadratic
action turns out to be more complicated, but still
the practical purpose of calculating the combinations of
chiral determinants entering in
multiloop string amplitudes has been achieved
\ref\mloop{
\npb{E. Verlinde and H.  Verlinde}{288}{1987}{357};
\npb{D.  Lebedev and A.  Morozov}{302}{1986}{163};
\plb{E. Gava, R. Iengo and G. Sotkov}{207}{1988}{283};
}.
The above programme was stopped by the difficulty of dealing with
generic curves outside the class of the
$Z_N$ symmetric ones. On the other hand, already at genus 3, the
hyperelliptic curves represent a subvariety of codimension 1, i.e. a
subset of measure zero in the
moduli space. Also the interesting approaches to treat CFT's on general RS
proposed in refs.
\ref\berad{
\plb{M. Bershadsky and A. Radul}{193}{1987}{213}
}
\ref\knizn{V. G. Knizhnik,
{\it Sov. Phys. Usp.} {\bf 32} (11) (1989) (945).}
have been abandoned. At the present, one can only speculate if it is
possible to encode all the information about the perturbative
expansion of string theory
in the limit $N\rightarrow\infty$ of some CFT on the sphere
(see formula (12.14) in \knizn ).
In recent years some progress
in dealing with CFT on general RS can nevertheless
be reported. In particular,
an operator formalism for CFT's on general
RS has been constructed in refs
\ref\zn{\ijmpa{F. Ferrari and J. Sobczyk}{11}{1996}{2213}}
\ref\gen{\jgp{F. Ferrari and J. Sobczyk}{19}{1996}{287}}
and minimal models of CFT has been defined on $Z_N$ symmetric
curves
\ref\ae{S. A. Apikyan and C. J. Efthimiou, {\it Minimal Models of CFT
on $Z_N$ Surfaces}, hep-th 9610051.}
(before only the hyperelliptic case was studied in this context
\ref\crnk{\plb{C. Crnkovic, G.M. Sotkov and M. Stanishkov}{220}{1989}{397}}
).

In this paper a modified approach to the
problem of constructing CFT's on RS
is put forward so that the global geometrical aspects can be
taken into account. One obtains a new algebraic structure (infinite
dimensional Lie algebra) which
should be viewed as an analogue of the Virasoro algebra. It is not
the first time that such structure is proposed
\ref\kn{
I. M. Krichever and S. P. Novikov, {\it
Funk. Anal. Pril.} {\bf 21} No.2 (1987), 46;
{\bf 21} No.4 (1988), 47;
J. Geom. Phys. {\bf 5} (1988) 631.
}. The algebra proposed in this paper
is very different but there seems to be a
common spirit which underlies both constructions.
Our construction can be viewed as a kind of
multipoint generalization of Krichever-Novikov basis: there are (provided
they are not branch points) exactly $N$ points (one on each sheet) with
coordinate $z=0$ and $z=\infty$. The
elements of the basis in which the solutions of the classical equations of
motion are expanded, are allowed to have poles only at
these points
\ref\dick{\fph{R. Dick}{40}{1992}{519}.
}.

Technically, the difficulties to deal with CFT on generic RS
are related with their monodromy group. In the $Z_N$ symmetric case all the
monodromy matrices are ``the same'' due to the peculiar symmetry
of the RS and can be simultaneously diagonalized. For
general RS this is impossible. Take for example a generic curve of
genus $g=3$. This can be viweved as a three-sheeted covering of the complex
sphere with ten branch points. 
Two sheets (for instance the second and third) are glued together
at two branch points
while the first and second sheets are glued at the
remaining branch points.
The monodromy matrices associated to the first two
branch points are $\pmatrix{1&0&0\cr 0&0&1\cr 0&1&0}$ and the
remaining eight are $\pmatrix{0&1&0\cr 1&0&0\cr 0&0&1}$. Clearly, they do not
commute and so they cannot be simultaneously diagonalized.
But something else can be done:
any analytic tensor field
of spin $\lambda$
(also called here
meromorphic $\lambda-$differential) with given poles and zeros
can be decomposed into a sum of independent
components modulo single-valued functions with different
monodromy
properties. These components, whose number is equal to the
number of sheets of the RS in question, can be explicitly constructed.
Of course, there are many possible
monodromy decompositions of this kind.
>From the point of view of the construction presented in this paper, all of
them seems to be equivalent, but to make computations easier
it is important to find the simplest possible decomposition. This
point will be discussed in Chapter 5 in more details.

Assume that the RS is given by means of algebraic equation 
(polynomial) in two complex
variables $z$ and $y$:

\eqn\aleq{ F(z,y) = y^NP_N(z) + y^{N-1}P_{N-1}(z) + ... + P_0(z) = 0 }

where $P_j(z)$ are polynomials in $z$. $y$ is a single-valued
function on the RS but one can also consider
$y(z)$ as multivalued function on the complex sphere. The monodromy
properties of $y(z)$ define the RS. We shall adopt the notation
$y^{(j)}(z)$, where $j=1,2,...,N$ refers to the value of $y$ on the $j$
branch of the
RS. $z$ plays a double role in the construction. It can be viewed either as
single-valued function on the RS
or as a coordinate (with trivial monodromy properties) on the complex
sphere on which the $N$ sheets composing the RS are
projected.
Transporting the function $y(z)$ along a small closed path around a given
branch
point located at $z=a$,
its branches are exchanged according
to the local monodromy properties of the RS in $a$.
Denoting with
$m_a[f]$ the operator which transport a multivalued
function $f(z)$ around $a$,
it is easy to see that:

\eqn\mon{ m_a \left( y^{(j)}(z)\right) = \sum_{k=1}^N
M_{(a)}^{jk} y^{(k)} (z)}

where the $M_{(a)}^{jk}$ are $N$-dimensional permutation matrices.
All the meromorphic tensor fields can be constructed in
terms of powers of $dz$ and rational functions of $z$ and $y$ \grha .
It is clear that the nontrivial monodromy is carried only by $y$.
The simplest monodromy decomposition of a tensor field $\omega$
being a meromorphic $\lambda$-differential is given by

\eqn\modec{ \omega (z) = \sum_{k=0}^{N-1} g_k(z) y^k(z) dz^{\lambda}
}

where $g_k(z)$ are rational functions of $z$. An elementary proof of that fact
can be found in \gen .

The monodromy decomposition is a basic idea behind the construction of
the operator formalism on RS in \zn - \gen .
Usually by CFT's on RS one understands a set of
correlation functions satisfying necessary analytic properties.
These "physical" requirements are typically
strong enough
to define correlators up to an overall normalization
\ref\raina{\hpa{A. K. Raina}{63}{1990}{694};
{\it Comm. Math. Phys.}, {\bf 140} (1991), 373 .
}.

In all approaches to establish an operator formalism for CFT's on RS,
one tries to construct
a
vacuum state and to act on it with annihilation and creation operators as in
ordinary quantum field theories in flat spaces, see e. g.
\ref\altern{
\cmp{L. Bonora, A. Lugo, M. Matone and J. Russo}
{123}{1989}{329};
\plb{C. Vafa}{190}{1987}{47};
\npb{L. Alvarez-Gaum{\' e}, C. Gomez, G. Moore and C. Vafa}{303}{1988}
{455};
\plb{A. M. Semikhatov}{212}{1988}{357}.
}.
In the approach of \gen,  the $b-c$ fields are represented in a suitable way
in monodromy decomposition form.
Two different expansions are chosen for $b$ and $c$. In this way the
the simplest possible anticommutation
relations between the elementary excitations of both fields can be postulated.

On arbitrary genus RS defined by means of \aleq\ such decompositions
are given by:
\eqn\bj{
b(z) = \sum_{k=0}^{N-1}\sum_{i=-\infty}^{+\infty}
b_{k,i}z^{-i-\lambda}f_k(z)
}

\eqn\cj{ c(z) = \sum_{k=0}^{N-1}\sum_{i=-\infty}^{+\infty}
c_{k,i}z^{-i+\lambda -1} \phi_k(z)
}

The nontrivial monodromy is carried by the (multivalued) functions

\eqn\dfk{
f_k(z) = {y^{N-1-k}(z)dz^{\lambda}\over (F_y(z, y(z)))^{\lambda}}
}

and

\eqn\dfik{
\phi_k(z) = {dw^{1-\lambda}\over (F_y(z, y(z)))^{1-\lambda}}
\left(y^k(z) + y^{k-1}(z)P_{N-1}(z) + ... + P_{N-k}(z)\right) .
}
where $F_y\equiv {\partial F\over \partial y}$.
The single-valued functions multiplying them are powers of $z$ as in the
usual Laurent expansion.
Except for the monodromy contributions
we follow as close as possible the way in which
the CFT's
on the complex sphere are treated.
To go to the quantum theory we postulate the
following anticommutation relations

\eqn\elant{
\{ b_{s,j}, c_{r,k}\} = \delta_{s,r}\delta_{j+k,0}
}

After explaining in \gen\ all the details concerning the construction of
vacuum state, splitting of oscillators into annihilation, creation and
zero mode part, it is natural to continue the same analysis in the
case of energy momentum tensor.
This is an important object, because it generates the
conformal transformations and its vacuum expectation value contains
information about the dependence of the partition function of the theory
on the moduli.

The monodromy components of the energy momentum tensor for $b-c$ system can
be expressed in terms of the
elementary excitations $b_{k,i}$ and $c_{k,i}$
of the fields $b$ and $c$. Once they are known,
the commutators of the resulting  operators can  then be calculated.
One obtains an infinite dimensional Lie
algebra. More precisely, at this stage of the computations one disregards
central extension terms related to normal ordering ambiguities.
It will be shown that there is a
simpler way to include these terms.
It turns out that the algebra can be understood
in terms of an operator product expansion (OPE)
for the monodromy components of $T(z)$.
This is one of two most important results of this paper. In deriving it,
the usual techniques (radial quantization)
of the operator formalism
for CFT on the complex sphere are applied.
The language of OPE makes it possible
to calculate
also a central extension of the algebra.

In order to ilustrate in details the above ideas the simpler
hyperelliptic case is
studied in Chapter 2. The monodromy decomposition of $T(z)$ gives rise to
generators $L_{1,n}$ and $L_{2,m}$. Their Lie algebra brackets
are explicitly computed.
In the standard operator formalism
on the sphere the conjugation of conformal fields is
understood in the framework of the
"radial quantization", where $z=0$ corresponds to asymptotic "in" states and
$z=\infty$ to asymptotic "out" states. Time inversion is expressed as
$z\rightarrow {1\over \bar z}$.
In the case of the elementary excitations for the fields
$b$ and $c$ on RS our natural
construction of the representation space leads itself
to a definition of conjugation
\ref\mb{F. Ferrari and J. Sobczyk,
{\it Quantum Field Theories on Algebraic Curves}, to appear in the Proceedings
of the IX Max Born Symposium, September 1996, Karpacz, (PL).
}.
In Chapter 3 it is shown that this definition
is equal to the one suggested by
the radial quantization prescription.
One finds that also the expressions of the operators
$L_{1,j}$ and $L_{2,j}$ in terms of the multivalued modes
in which the fields
$b$ and $c$ have been expanded
are subjected to a conjugation consistent with radial quantization
considerations.
This is the second important
result of the paper.
A surprising feature of the conjugation is that it involves
parameters of the algebraic
equation defining RS. The origin of this dependence is clearly
explained.

One can try to reconstruct the generators $l_n$ of the Virasoro
algebra as infinite linear combination of $L_{1,n}$ and $L_{2,n}$.
In the hyperelliptic case such representation is given but
this procedure should be treated with caution. It is unclear if it can
be generalized to the case of arbitrary algebraic curves.
Also, the reconstructed $l_n$'s do not have the property
$l_n^{\dagger}=l_{-n}$.

The motivation of our investigations is the possibilty of finding
analogous structures associated
also to arbitrary RS. It is always possible
to find the monodromy decomposition of $T(z)$ and postulate
simple OPE for its monodromy components. There is some freedom
in the form of the OPE so that there are free numerical
parameters in the infinite dimensional Lie algebra. Performing in
$F(z,y)$ the transformation $z\rightarrow {1\over\bar z}$ it is
possible to introduce a
natural notion of conjugation. The hope is that thanks to the
study of the detailed structure of
this algebra one can get information
about the physical content of the theory. In the case of the Virasoro algebra,
the theory of its representations was a basic tool to investigate
structure of CFT \bpz .
It is important that one can always
perform explicit computations following those presented in Chapter 2 and
compare them with the structure produced by OPE.

The new algebraic structure should
be studied in more detail. One should address questions about
the definition of primary fields, the general properties of
representations, null states, unitary
representations etc. All these issues will be discussed in future
papers.

\vskip 1cm
\newsec{ALGEBRA IN THE HYPERELLIPTIC CASE}
\vskip 1cm

Let us consider the $b-c$ systems with action:

\eqn\action{S_{bc} ={1\over \pi} \int_{RS} d^2z \left( b\bar\partial c
+ {\rm c.c.}\right)}

where $z$ and $\bar z$ are complex coordinates on the RS,
$\bar\partial\equiv{\partial
\over \partial\bar z}$.
$b$ and $c$ are fermionic fields carrying conformal weights $\lambda$
and $1-\lambda$ with classical equations of motion

\eqn\cem{ \bar\partial b = \bar\partial c = 0.}

The energy momentum tensor is

\eqn\ener{
T(z) = -\lambda b(\partial c) +
(1-\lambda ) (\partial b) c}

Here we investigate the special case of
hyperelliptic curves defined by the
equation ($\infty$ is not a ramification point):

\eqn\heq{y^2 = \prod_{j=1}^{2g+2} (z-a_j).}

In accordance with \zn\ and \gen\ the following monodromy
decomposition of both fields are postulated:

\eqn\bexp{b(z)= \sum_{k=0}^1\sum_{j=-\infty}^{+\infty}
b_{k,j} {z^{-\lambda -j}\over y^{\lambda -k}} dz^{\lambda}
}

\eqn\cexp{c(z) = \sum_{k=0}^1
\sum_{j=-\infty}^{+\infty}
c_{k,j} {z^{\lambda -j-1}\over y^{k-\lambda}} dz^{1-\lambda}
.}

One recognizes that they represent particular cases of \bj\ and \cj.
The elementary excitations satisfy the relation (see \elant )

\eqn\flant{
\{ b_{s,j}, c_{r,k} \} = \delta_{s,r} \delta_{j+k,0}
}

The following notation will be useful:

\eqn\nota{y^2 = \sum_{j=0}^{2g+2} A(j) z^j.}

It will be understood that

\eqn\notb{A(j) = 0\qquad {\rm for}\qquad j>2g+2.}

Let us introduce the following monodromy decomposition of the
energy momentum tensor:

\eqn\enmom{T(z) = \sum_{k=1}^2 T_k(z) = \sum_{k=1}^2\sum_{j=-\infty}^{+\infty}
L_{k,j} {z^{-j-2}\over
y^k}dz^2.}

The reason to take $k=1,2$ rather then $k=0,1$ is merely technical. With the
adopted choice it is simpler to express $L_{k,j}$ in terms of
$b_{r,n}$ and $c_{s,m}$.

In the hyperelliptic case in which the monodromy group can be diagonalized
one should distinguish
between the above "monodromy decomposition" and the
treatment of $T(z)$ in \kniz, where the values of the energy
momentum tensor on two branches of the curve: $T^{(k)}$, $k=1,2$
are arranged into linear combinations which are diagonal with respect to the
monodromy group

\eqn\dcom{T^{\pm}(z) = {1\over \sqrt{2}} \left( T^1(z) \pm T^2(z) \right)
}

Thus in the notation explained in the introduction, the elements of the
monodromy matrices are:

\eqn\monel{ M^{++} = -M^{--} = 1,\qquad M^{+-} = M^{-+} = 0.}

Comparing eqs. \enmom\ and \dcom\ it seems possible
to identify $T^+$ with $T_2$ and
$T^-$ with $T_1$. This is because $T_2$ has trivial monodromy properties
($y^2$ can be expressed as polynomial in $z$). That correspondence is
however misleading.
By adding the components
of $T(z)$ in both approaches one obtains:

\eqn\comp{ T_1 + T_2 = T,\qquad T^+ + T^- = \sqrt{2} T^1.}

A correct way of thinking about $T^{\pm}$ is that of
fields on ${\bf C}P(1)$ while $T_k$ should rather be treated as fields on
RS.

The advantage of the representation \enmom\ is that it can be generalized
to the case of general algebraic curve when the
monodromy group cannot be globally diagonalized.

The energy momentum tensor \ener\ can be expressed in terms of the
$b$ and $c$
fields modes introduced in \bexp\ and \cexp . By looking at the monodromy
decomposition of $T(z)$ it is possible to express $L_{k,m}$ in terms
of $c_{s,n}$ and $b_{r,m}$:

\eqn\ela{L_{2,p} =  \sum_{k=0}^1
\sum_{m=-\infty}^{+\infty}\sum_{j=-\infty}^{+\infty}
q^{(2-k)}_{p,m,j} :b_{k,j} c_{k,m-j}:
}

\eqn\elb{L_{1,p} = \sum_{k=0}^1
\sum_{m=-\infty}^{+\infty}\sum_{j=-\infty}^{+\infty}
s^{(2-k)}_{p,m,j} :b_{k,j} c_{1-k,m-j}:
}

where

\eqn\elc{
q^{(2)}_{p,m,j} = A(m-p)\left({\lambda\over 2}(m+p) -j\right)
}

\eqn\eld{q^{(1)}_{p,m,j} = A(m-p) \left({\lambda\over 2}(p+m) + {1\over 2} (m-p)
-j \right) }

\eqn\ele{s^{(2)}_{p,m,j} = \delta_{p,m} (\lambda p-j)}

\eqn\elf{s^{(1)}_{p,m,j} = A(m-p)\left(\lambda p -j + {1\over 2}
(m-p)\right) }

The definition of normal ordering :\ : will be given in Chapter 4.
It follows from \elc\ - \elf\ that sums over $m$ in \ela\ and \elb\ are
finite (see \notb ).
It is possible to calculate the algebra commutators. At this stage
the full structure will not be obtained,
namely central
extension terms will be omitted.
There is much easier way to
calculate them.

Commutators of the algebra generators are:

\eqn\coma{[L_{2,p}, L_{2,r}] =
(p-r) \sum_{s=p+r}^{p+r+2g+2}
A(s-p-r) L_{2,s}\qquad {\rm + c.e.\ terms}}

\eqn\comb{[L_{1,n}, L_{1,m}] = (n-m) L_{2,n+m}\qquad {\rm + c.e.\ terms}}

\eqn\comc{[L_{2,p}, L_{1,n}] =
\sum_{s=n+p}^{n+p+2g+2} {1\over 2} (s + p - 3n) A(s-n-p) L_{1,s}\qquad
}

(in $L_{1,n}$ there is no normal ordering ambiguity). It will be shown
in Chapter 4 that the normal ordering involves only $L_{2,n}$ with
$n\leq 0$. It follows that central extension terms are present
only for $p+r\leq 0$ in \coma\ and $n+m\leq 0$ in \comb .

\vskip 1cm
\newsec{OPE FOR THE MONODROMY COMPONENTS OF $T(z)$}
\vskip 1cm

In the approaches of
\bera , \kniz , \knizn , \ae , \crnk\
the OPE for diagonal (with
respect to monodromy) components of $T(z)$ are postulated.
They have the property to keep the monodromy, i.e. the
monodromy properties (which are described by single numbers,
"charges") of
both sides of OPE are the same. This can be adopted as a
principle according to which
more general situations are handled.
For this purpose the identification of
components of $T(z)$ in the way
explained in \enmom\ is very useful.

The requirement of unchanged monodromy plus selfconsistency of the whole
structure
leads to the following OPE's
($\Gamma ,\Gamma ', D, E, F, G, H$ are constants to be fixed):

\eqn\mona{T_2(z) T_2(w) \sim {\Gamma '\over (z-w)^4} + {2DT_2(w)\over (z-w)^2}
+ {D\partial T_2(w)\over z-w}}

\eqn\monb{T_1(z) T_1(w) \sim {\Gamma\over (z-w)^4} + {2HT_2(w)\over (z-w)^2}
+ {H\partial T_2(w)\over z-w}}

\eqn\monc{T_2(z) T_1(w) \sim {2ET_1(w)\over (z-w)^2}
+ {F\partial T_1(w)\over z-w}}

\eqn\mond{T_1(z) T_2(w) \sim {2ET_1(w)\over (z-w)^2}
+ {G\partial T_1(w)\over z-w}}

In the above OPE's the symmetry properties under the transformation
$z\leftrightarrow w$ have
been used. For example, in \mona\ one could have started with

\eqn\monaa{ T_2(z) T_2(w) \sim {\Gamma '\over (z-w)^4} + {2DT_2(w)\over (z-w)^2}
+ {\tilde D\partial T_2(w)\over z-w}}

but a simple analysis leads to condition $D = \tilde D$.

The standard OPE for $T(z) = T_1(z) + T_2(z)$ implies

\eqn\kons{D+H = 2E = F + G = 1}

and

\eqn\konsa{\Gamma + \Gamma ' = {c\over 2} = - (6\lambda^2 -6\lambda +1)}

where $c$ is the conformal anomaly of CFT.
The operators $\tilde L_{k,n}$ can be expressed in terms of $T(z)$ via contour
integrals. We find it useful to distinguish here $L_{k,n}$ expressed in
terms of elementary excitations $b_{k,n}$ and $c_{k,n}$ from $\tilde
L_{k,n}$ as, a priori, they need not satisfy the same algebra:

\eqn\aaa{\tilde L_{k,j} = {1\over \pi i}\oint T_k y^k z^{j+1} dz.}

It is crucial to notice that the integral \aaa\ is well defined even if
$T_k$ takes different values on different branches of the RS. The
combination $T_k y^k$ (no summation in the index $k$!) is well defined on ${\bf
C}P(1)$ as it has trivial monodromy properties. The origin of unusual
normalization can be traced back in \comp .

Using standard techniques of CFT (radial quantization etc) one obtains

$$[\tilde L_{2,p}, \tilde L_{2,r}] =
2D (p-r) \sum_{s=p+r}^{p+r+2g+2} A(s-p-r) \tilde L_{2,s} +$$
\eqn\ccoma{
+ {2\Gamma '\over 3}\sum_{s={\rm max}(r,-p-2g-2)}^{{\rm min}(-p,2g+2+r)}
A(-r+s)A(-p-s)\big(
(s+p-r)^3 - (s+p-r) \big) .
}

A central extension term is present if $-(4g+4)\leq p+r\leq 0$.

$$[\tilde L_{1,n}, \tilde L_{1,m}] = 2H (n-m) \tilde L_{2,n+m}
+ {2\Gamma\over 3} \Bigg( {1\over 2}A(-n-m)(n^3-m^3+m-n) +$$

\eqn\ccomb{
- {3\over 8}
\Big( \sum_{j=1}^{{\rm min}(2g+2,-n-m)}
\ \ \sum_{k=1}^{{\rm min}(2g+2,-n-m-j)}
(n-m)jkA(j)A(k)B(-n-m-j-k) \Big)\Bigg)
}

Here again a central extension term is present for $-(4g+4)\leq n+m\leq 0$.
Finally we have:
\eqn\ccomc{[\tilde L_{2,p}, \tilde L_{1,n}] =
\sum_{s=n+p}^{n+p+2g+2}
\Big( s(4E-3F) + pF + n(F-4E)\Big) A(s-n-p) \tilde L_{1,s}.
}

In the above  formulae the $B(s)$'s,
 ($s\geq 0$), are defined by means of the relations:

\eqn\iden{\sum_{n={\rm max}(0,m-2g-2)}^{m\geq 0} B(n) A(m-n) = \delta_{m,0}.}

for $s'<0$ $B(s')=0$.
In addition to central extension terms the generators $\tilde
L_{k,j}$ satisfy the same algebra as $L_{k,j}$ provided that
$D=H=F=E={1\over 2}$. It is interesting to notice that the two numerical
constants $\Gamma '$ and $\Gamma$ are not independent. The Jacobi
identities for the whole
structure imply

\eqn\conce{ \Gamma '= \Gamma . }

In the explicit verification of Jacobi identities
one has to use the identity

\eqn\idb{ \sum_{s=0}^w \Big(
2(\gamma -2\delta )s^3 - 3(\gamma -2\delta ) s^2 w
+ \gamma s w^2 - \delta w^3 \Big) A(s) A(-s+w) = 0}

where $w$ is an arbitrary integer $w\geq 0$. $\delta$ and $\gamma$ are
arbitrary integers.

One can try to identify the generators of the Virasoro algebra $Vir$.
The obvious strategy is to
expand
$T(z)$ in a power series around $z=0$. It results in

\eqn\femt{T(z) = \sum_{n=-\infty}^{+\infty}
l_n z^{-n-2}}

with

$$
l_n = {1\over 2}\Big( (A(0))^{-{1\over 2}} L_{1,n} - {1\over 2}
(A(0))^{-{3\over 2}} A(1) L_{1,n+1} + ... +
$$
\eqn\tva{
+ A(0)^{-1}L_{2,n} - (A(0))^{-2} A(1) L_{2,n+1} + ...\Big)
}

The argument leading to \tva\ is rather delicate.
It is necessary to
expand ${1\over y}$
around $z=0$ and it is unclear which branch of $y(z)$ one
should take. Actually,
in the hyperelliptic case both values differ only by sign.
The generalization of \tva\ to the case of generic RS is rather
problematic.

The
perturbative computations confirm that the $l_n$ satisfy
the algebra

\eqn\tvb{
[l_n, l_m] = (n-m)l_{n+m} + {c\over 12}(n^3 - n)
}

with

\eqn\tvc{ c={\Gamma\over 4} = {\Gamma '\over 4}.
}

The algebra \ccoma -\ccomc\ turns
out to be symmetric under the transformations

\eqn\syma{ L_{2,n} \rightarrow L_{2,-n-(2g+2)} }

\eqn\symb{ L_{1,n} \rightarrow L_{1,-n-(g+1)} }

\eqn\symc{ A(j) \rightarrow A(2g+2-j) }

The term containing $B$ in \ccomb\ should be subject to
transformation which follows from the definition \iden :

\eqn\symd{ B(j) \rightarrow B(-2g-2-j). }

The symmetry \syma -\symc\ can be verified by direct inspection. In the next
Chapter it will be explained why it is present.

\vskip 1cm
\newsec{EXPLICIT REALIZATION}
\vskip 1cm

In the beginning of this Chapter general RS are considered.
The operator formalism of \gen\ gives rise to a natural notion of
normal ordering. The operators are either
creation and annihilation or they correspond to zero modes
(there are $M_s$ such operators $b_{s,j}$ for each value of $s$).
Obvious normal ordering rules have to be supplemented
requiring that
zero mode excitations should be treated as creation operators.

The following oscillators are annihilation operators:

\eqn\opan{b_{k,j}|0> = 0, \qquad j\geq 1-\lambda,\qquad
c_{k,j'}|0> = 0, \qquad j'\geq \lambda}

A vacuum state $|0>$ can be constructed with the properties

\eqn\vacdefa{ <0|0>=0 }

\eqn\vacdefb{ <0|\prod_{s=0}^{N-1}\prod^{-\lambda}_{j = 1-\lambda -M_s}
b_{s,j} |0> = 1}

The fact that the elementary excitations satisfy anticommutation rather then
commutation relations implies that $|0>$ should be of the form
of "Dirac sea" or, in other words, a semi-infinite wedge product:

\eqn\vacdefc{ |0> = \prod_{k=0}^{N-1} \beta_{k,1-\lambda}\wedge
\beta_{k,2-\lambda}\wedge \beta_{k,3-\lambda}\wedge ... }

It will be assumed that "sectors" of the theory labeled by different indices
$k$ commute.
Elementary oscillators should act on vectors from
the representation space according to:

\eqn\alrel{ b_{k,j} \hookrightarrow \beta_{k,j}\wedge...\qquad
c_{k,j} \hookrightarrow {\partial\over \partial \beta_{k,-j} }.}

A bilinear form reproducing
\vacdefa -\vacdefb\ is given in
the following way. For

\eqn\stx{ |x> = \prod_{k=0}^{N-1}
\beta_{k,j_{k_1}}\wedge \beta_{k,j_{k_2}}\wedge ...}

\eqn\stx{ |y> = \prod_{k=0}^{N-1}
\beta_{k,j_{k_1}'}\wedge \beta_{k,j_{k_2}'}\wedge ...}

one defines

\eqn\scpr{ <y|x> \equiv \prod_{k=0}^{N-1} ...\wedge\beta_{k,-j_{k_2}'+H_k}
\wedge\beta_{k,-j_{k_1}'+H_k}\wedge
\beta_{k,j_{k_1}}\wedge \beta_{k,j_{k_2}}\wedge ... }

where $H_k=1-2\lambda -M_k$. By definition, \scpr\ is equal zero
unless all the excitations are present there. If not zero it is
equal $1$.

This choice of the bilinear form implies

\eqn\bcon{b^{\dagger}_{k,j} = b_{k,-j+H_k}\qquad c^{\dagger}_{k,j}
= c_{k,-j-H_k}. }

This definition of conjugation may seem unnatural, but
there is a clear explanation for that.

Let us consider from now on the hyperelliptic case. The numbers of $b$ zero
modes for $k=0,1$ are equal to

\eqn\nzm{ M_0 = \lambda (g+1) - 2\lambda + 1,\qquad M_1 =
\lambda (g+1) - 2\lambda - g }

so that

\eqn\vh{ H_0 = - \lambda (g+1),\qquad H_1 = -(\lambda - 1)(g+1).}

The meaning of normal ordering in \ela\ and \elb\ is now clear.
>From the definition of $L_{k,n}$ it is obvious that some of them are
free from normal ordering ambiguities. This is true for
all $L_{1,j}$ and also for
$L_{2,j}$ with $j\geq 1$. Automatically, the same applies to $l_j$ (generators
of Virasoro algebra - see \tva ) for $j\geq 1$.

The operators $L_{2,j}$ for $j\geq 1$
annihilate the vacuum state. One can choose the normal ordering constant
in $L_{2,0}$ in such a way that it also
annihilates vacuum

$$:L_{2,0}: = \sum_{k=0}^1 \Big( \sum_{m=1}^{2g+2}\sum_{j\leq m-\lambda}
A(m)\left( {(\lambda +k)m\over 2} - j\right) b_{k,j}c_{k,m-j} +$$

$$
- \sum_{m=1}^{2g+2}\sum_{j >m-\lambda}
A(m)\left( {(\lambda +k) m\over 2} - j\right) c_{k,m-j} b_{k,j} \Big) +$$

\eqn\opani{
+ A(0)\sum_{k=0}^1\left(
\sum_{j\geq 1-\lambda} j c_{k,-j}b_{k,j}
- \sum_{j\leq -\lambda} j b_{k,j}c_{k,-j}\right) .
}

The normal ordering for the generators
$L_{1,j}$ is not modified. The $L_{1,k}$'s
for $k\geq 0$ annihilate vacuum state.
In the standard operator formalism for CFT's on the complex sphere the
definitions of bilinear form and conjugation are closely related with
radial quantization. The complex sphere is equipped with an Euclidean metric
and the time involution acts on the "space-time" coordinate $z$ according to
$z\rightarrow {1\over \bar z}$. There is a $1-1$ correspondence between
elements of the Hilbert space and the conformal fields

\eqn\hscft{
|\psi_{\rm in}> = \lim_{z,\bar z \rightarrow 0} \psi (z,\bar
z) |0>.
}

A natural definition of conjugation is

\eqn\condf{\psi^{\dagger}(z,\bar z) = \bar z^{-2\lambda} z^{-2\bar\lambda}\psi
({1\over\bar z},{1\over z}.)
}

It is assumed that $\psi$ is a conformal field of dimensions
$\lambda$ and $\bar\lambda$.

Applying the above rules also to the fields $b-c$
one obtains e.g.

\eqn\geint{
\sum_{j=-\infty}^{+\infty} b_{0,j}^{\dagger}{\bar z^{-\lambda-j}\over
y^{\lambda}(\bar z)}
= \sum_{j=-\infty}^{+\infty} b_{0,j} \bar
z^{-2\lambda}
{\bar z^{\lambda +j}\over y^{\lambda} ({1\over\bar z}) }
}

It is very useful to define $\tilde y$ in the following manner

\eqn\tydf{
y^2({1\over z}) = \sum_{j=0}^{2g+2} A(j)z^{-j} =
z^{-2g-2}\sum_{j=0}^{2g+2} \tilde A(j) z^j = z^{-2g-2}\tilde y^2 (z)
}

with

\eqn\tydtb{
\tilde A(j) = A(2g+2 -j).
}

With that definition one can write

\eqn\guint{
\sum_{j=-\infty}^{+\infty} b_{0,j}^{\dagger}{\bar z^{-\lambda-j}\over
y^{\lambda}(\bar z)} = \sum_{j=-\infty}^{+\infty} b_{0,j} \bar
{\bar z^{-\lambda + j + \lambda (g+1)}\over \tilde y^{\lambda}(\bar z)}
}

It is natural
to define $b_{0,j}^{\dagger} =
b_{0,-j-\lambda (g+1)}$.
This conjugation has in general to be taken together with the transformation
$A\rightarrow \tilde A$ \tydtb .
The definition of conjugation for $b_{k,j}$ and $c_{k,j}$ works
perfectly well since the algebra \flant\ does not contain $A(j)$.
On the other
hand the algebra of $L_{k,j}$ includes explicitly $A(j)$ and so the
conjugation for $L_{k,j}$ 

\eqn\conl{ L_{1,j}^{\dagger} = L_{1,-j-(g+1)},\qquad L_{2,j}^{\dagger}
= L_{2,-j-(2g+2)}.}

(eq. \conl can be deduced in the same way as
\guint ) has to be supplemented by the transformation
$A(j)\rightarrow \tilde
A(j)$.

\vskip 1cm
\newsec{POSSIBLE GENERALIZATIONS}
\vskip 1cm

The motivation to develop the monodromy
formalism is a hope to apply it to a generic RS given by
\aleq .
Let us describe in general terms what the features of such
construction should be.

One can use the operator formalism for
$b-c$ system and the  elementary excitations to define $L_{k,j}$ as it
was done in Chapter 2. There are several choices for the monodromy
decomposition of $T(z)$. One possibility is

\eqn\ansa{T(z) = \sum_{k=0}^{N-1}\sum_{j=-\infty}^{+\infty}
L_{k,j}{z^{-j-2}\over y^k}dz^2}

The other possibility is suggested by the form of expansions for $b$ field
\bj\

\eqn\ansb{T(z) = \sum_{k=0}^{N-1}\sum_{j=-\infty}^{+\infty}
L_{k,j} {z^{-j-2} y^{2-k}\over (F_y)^2}dz^2.}

Whatever the monodromy decomposition is, one gets concrete
expressions for $L_{k,j}$ and
the whole algebra can be calculated.
Its general features can be read from OPE.
Let us concentrate on the case \ansa .
The requirement that the OPE keeps the monodromy properties of the
components of $T(z)$ implies a general structure which can be written
in symbolic notation. Let us present more detailed formulas in the case of
a genus three RS described by the algebraic equation:

\eqn\genthree{ y^3 + 3y P(z) - 2 Q(z) = 0.}

$P(z)$ is a polynomial
of degree $3$ and $Q(z)$ a polynomial of degree $4$. Monodromy implies

$$T_0 T_0 \sim T_0\ +\ {\rm c.e.\ term}$$
$$T_0 T_1 \sim T_1$$
$$T_0 T_2 \sim T_2$$
$$T_1 T_1 \sim T_2$$
$$T_1 T_2 \sim T_0 + T_2\ +\ {\rm c.e.\ term}$$
\eqn\mondec{T_2 T_2 \sim T_0 + T_1 + T_2\ +\ {\rm c.e.\ term}}

The meaning of the above symbolic notation is that for example

\eqn\gemon{
T_0(z)T_1(w) \sim {\alpha T_1(w)\over (z-w)^2}
+ {\beta \partial T_1(w)\over z-w}
}

where $\alpha$ and $\beta$ are constants.

The last two relations in \mondec\ follow from the identities:

\eqn\deca{{1\over y^3} = {1\over 2Q(z)} + {1\over y^2} {3P(z)\over 2 Q(z)}}
\eqn\decb{{1\over y^4} = {3 P(z)\over 4 Q^2(z)} +
{1\over y} {1\over 2 Q(z)} + {1\over y^2} {9P^2(z)\over 4 Q^2(z)}.}

The generators $L_{k,j}$ are expressible in terms of $T_k(z)$ by means
of loop integrals (before fixing the constants $\alpha$, $\beta$ etc. it is
better to call them $\tilde L_{k,j}$)

\eqn\genex{
\tilde L_{k,j} = {1\over 2\pi i} \oint T_k y^kz^{j+1}.
}

The integral \genex\ is well defined as the monodromy property of
the product of $T_k y^k$ (no summation) are trivial.
From
\mondec\ the commutation relations among the  $L_{k,j}$'s can be computed.

The conjugation $L_{k,j}^{\dagger}$ can be calculated either
directly in terms of
$b_{k,j}^{\dagger}$ and $c_{k,j}^{\dagger}$ or by applying directly to
\ansa\ the radial quantization techniques. It is clear that
conjugation has to act also on the parameters of the algebraic
equation.
The simple transformation $A(j)\rightarrow A(2g+2-j)$ obtained for
hyperelliptic curve will be replaced
by a more complicated one.

One should analyze in detail e.g. genus $3$ RS. Emphasis has to be be put
on the choice of the simplest possible monodromy decomposition of $T(z)$,
perhaps one
analogous to the $Z_3$ symmetric case.

\eqn\ansc{T(z) = \sum_{k=0}^{2}\sum_{j=-\infty}^{+\infty}
L_{k,j} z^{-j-2} \left( S(z, y(z))\right)^kdz^2}

where $S(z, y(z))$, which has nontrivial monodromy,
satisfies the equation $S^3 = W(z)$, $W(z)$ being rational function of $z$
(i.e. $S^3$ has trivial monodromy).


\listrefs
\bye